# Evaluating and improving the predictive accuracy of mixing enthalpies and volumes in disordered alloys from universal pre-trained machine learning potentials

Luis Casillas-Trujillo, Abhijith S. Parackal, Rickard Armiento, Björn Alling

Department of Physics, Chemistry and Biology (IFM), Linköping University, 58183 Linköping, Sweden.

**Abstract**

The advent of machine learning in materials science opens the way for exciting and ambitious simulations of large systems and long time scales with the accuracy of *ab-initio* calculations. Recently, several pre-trained universal machine learned interatomic potentials (UPMLIPs) have been published, i.e., potentials distributed with a single set of weights trained to target systems across a very wide range of chemistries and atomic arrangements. These potentials raise the hope of reducing the computational cost and methodological complexity of performing simulations compared to models that require for-purpose training. However, the application of these models needs critical evaluation to assess their usability across material types and properties. In this work, we investigate the application of the following UPMLIPs: MACE, CHGNET, and M3GNET to the context of alloy theory. We calculate the mixing enthalpies and volumes of 21 binary alloy systems and compare the results with DFT calculations to assess the performance of these potentials over different properties and types of materials. We find that small relative energies necessary to correctly predict mixing energies are generally not reproduced by these methods with sufficient accuracy to describe correct mixing behaviors. However, the performance can be significantly improved by supplementing the training data with relevant training data. The potentials can also be used to partially accelerate these calculations by replacing the *ab-initio* structural relaxation step.

**Introduction**

Currently we are witnessing an explosion of scientific works that develop and apply machine learning (ML) methods in materials science, driven by the increased availability of large datasets, improvement in algorithms, and growth in computing power. Machine learning can substantially reduce computational effort and it has already proven capable of considerably speeding up both fundamental and applied research. Accurate density functional theory (DFT) calculations typically require large resources and are limited to relatively small supercells, usually fewer than a thousand atoms, and a few hundred picoseconds of simulation time. In contrast, traditional empirical interatomic potentials (i.e., those not based on ML) are computationally extremely efficient, even for large-scale systems containing millions of atoms and macroscopic simulation times. These potentials enable simulations of complex processes, such as nanoindentation or crack nucleation. Recently, machine learning interatomic potentials (MLIPs) have emerged as a new class of potentials that use ML from *ab-initio* calculations as training data to construct the potentials [1,2]. Contrary to classical potentials, MLIPs do not require any empirical information or human intuition and can overcome the loss of accuracy by optimization of the potential function parameters through a machine learning algorithm. Hence, MLIPs significantly reduce computational costs while maintaining similar accuracy as the DFT data used for training [3,4]. It opens the possibility of exciting and previously inaccessible simulations, which may be able to tackle the elusive mesoscale with near DFT accuracy. These techniques could enable, e.g., the study of deformation and fracture mechanisms driven by 2D and 3D defects through molecular dynamics simulations of millions of atoms. While the speed and applicability of these methods makes them attractive and powerful, one needs to apply them with careful consideration of the limit of accuracy set by the quality and scope of their training data. In particular, a key question is to what degree these models perform well when they have to extrapolate outside the systems present in the training data, and how the accuracy of such extrapolation differs between materials systems and physical properties.

Universal pre-trained machine learning interatomic potentials (UPMLIPs) have recently emerged as an approach to simulate any system across the periodic table [5]. In this work, we evaluate the performance of three such potentials:M3GNET [6], CHGNET [7], and MACE [8] for one of the most technologically important areas of solid-state physics: alloy theory. All of them present impressive results for many different complex problems, including phase transformations, catalysis, and aqueous systems [6-8]. The generally impressive results may tempt researchers to apply these models as is, even though the original publications emphasize the need for careful evaluation of their accuracy and a possible need to fine-tune them for particular purposes and classes of systems.

In this work, we evaluate UPMLIPs in the context of alloy theory and calculate the mixing enthalpy at ambient pressure for 21 different isostructural binary alloy systems. The mixing enthalpy of isostructural binary alloys is one of the least extrapolating cases of energy predictions, as these systems are very close to the pure elements, which are well represented in the training data. The calculations of mixing enthalpies are also a common first step in any theoretical study of alloy thermodynamics, and a problem where it would be attractive to use methods which greatly reduce the computational expense. Mixing enthalpies are obtained as a difference between fairly close energies. It is a common expectation on calculations using physics-based models, such as DFT, to be far more accurate for such energy differences than indicated by the general mean absolute error, in particular for relative energies between systems with fairly consistent bonding physics--it is ensured by a significant systematic component of the errors of these models. However, one cannot be sure this behavior is reproduced by ML methods obtained by very particular optimized to minimize the general errors of the model. In our tests presented below, we see how this expectation generally does not hold for the ML potentials; they often do not even reproduce the correct sign for mixing enthalpies. However, we also show that UPMLIPs can still be used to accelerate these types of calculations by using it only for structural relaxation, which is then followed by a DFT calculation for the energy.

**Potentials**

All three potentials used in this paper were trained on data available from Materials Project [9], which contains crystal structures that are typically seeded from experimental inorganic structure databases such as ICSD [10]. DFT calculations in the Materials Project database are performed using the PBE-GGA or the GGA+U [11] methods, using consistent settings. To train the potentials, energy, force, and stress values are appropriately sampled from the relaxation trajectories in the Materials Project dataset, thereby capturing the DFT potential energy landscape around energy minimums [12].

The M3GNET is a graph neural-network based interatomic potential that uses the 3-body interactions from the crystal structures as features [6]. It was trained on a dataset from Materials Project, named MPF.2021.2.8, comprising 187,687 ionic steps with energies, a total of 16,875,138 force components and 1,689,183 stress components from 62,783 compounds spanning 89 elements in the periodic table. As far as we have been able to conclude from the dataset, the pretrained model is using energies that have not been modified by the framework of compatibility corrections often applied in the context of data from Materials Project [13,14]. The model was reported to give a mean absolute error (MAE) of 0.01 eV per atom for energies, 0.033 eV Å$^{-1}$ for forces, and 0.042 GPa for stresses on a validation dataset.

The CHGNET potential is also based on a graph neural network architecture, which additionally includes the on-site magnetic moments to enable charge-informed atomistic modelling [7]. It is trained on the Materials Project Trajectory (MPtrj) dataset [15] containing 1 580 395 relaxation trajectory snapshots of structures with energies, 7 944 833 magmoms, 49 295 660 forces, and 14 223 555 stresses. In contrast to M3GNET, the energies used in training have been adjusted using the compatibility corrections mentioned above, including the GGA/GGA + U mixing correction. The model was reported to give a MAE of 0.03 eV per atom for energies, 0.07 eV Å$^{-1}$ for forces, a 0.348 GPa in stresses, and a 0.032 μB for magnetic moments on a validation dataset.

The MACE potential is an equivariant message passing graph tensor network that uses high body order equivariant features using Atomic Cluster Expansion [16,17]. The pre-trained uses 4-body interactions and was also trained on the MPtrj dataset, (however, in this case without the compatibility corrections). The potential was published along with tests to demonstrate its versatility in applications ranging from predicting phonon spectra to simulating solvent mixtures [8].

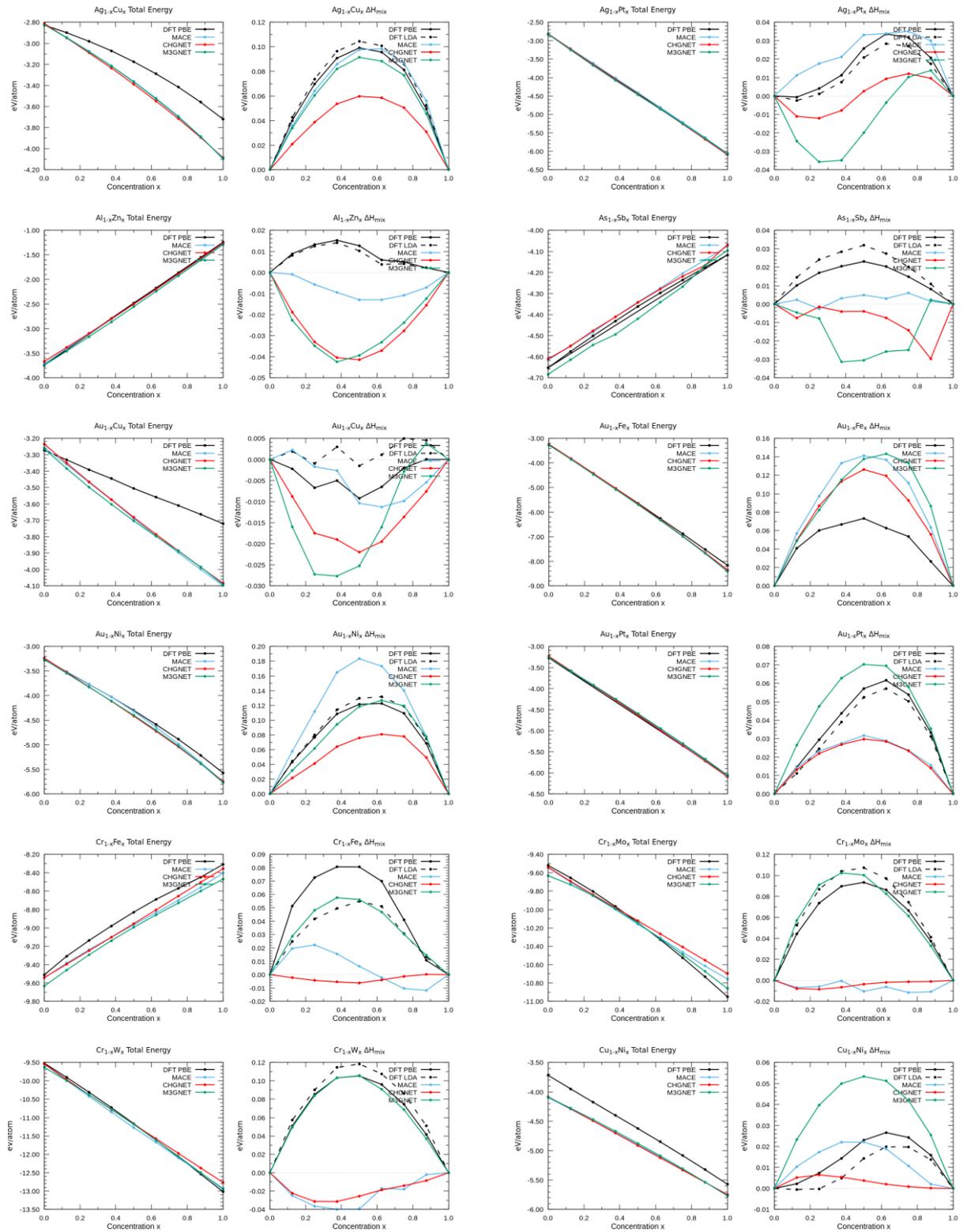

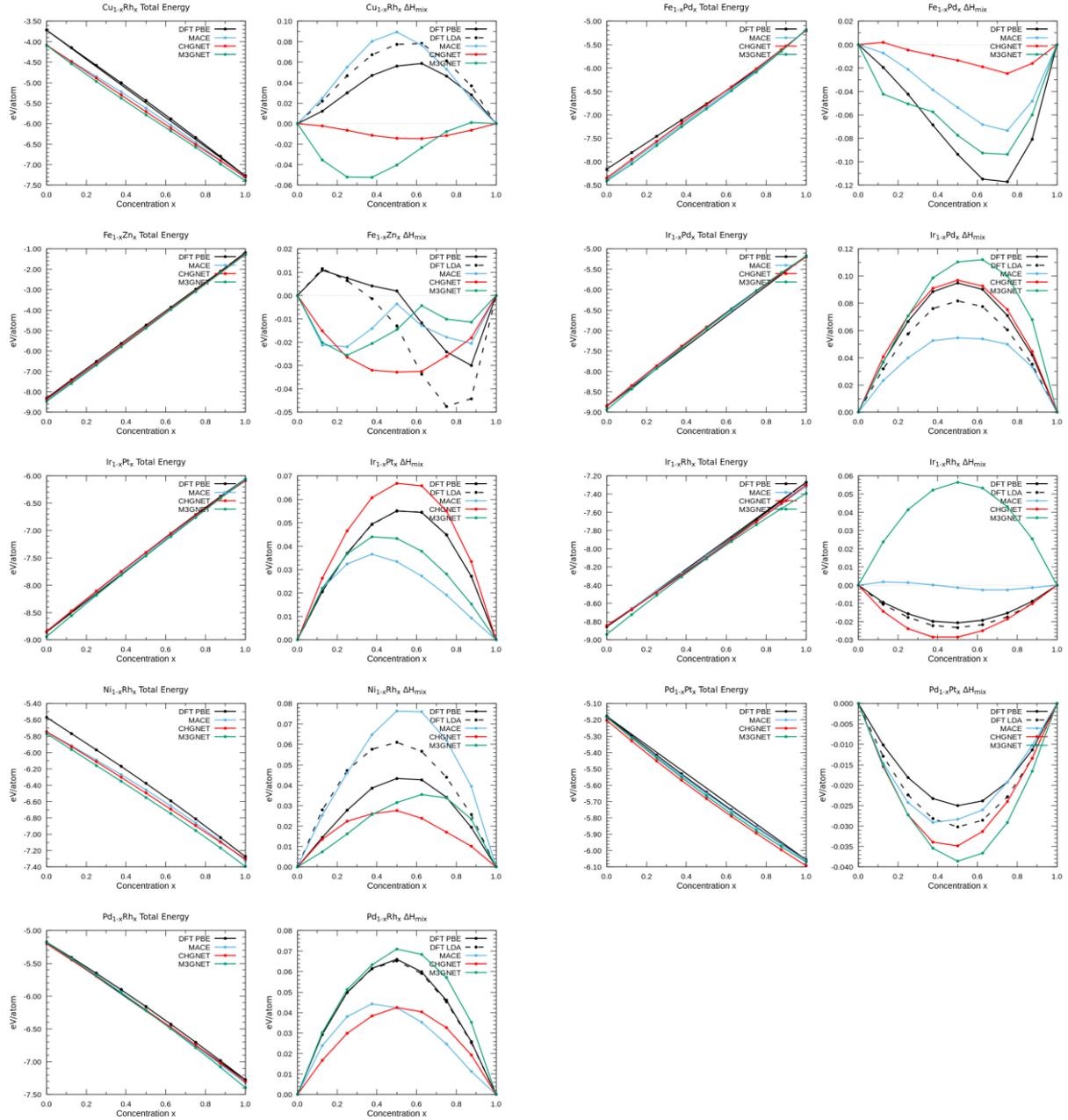

Figure 1. Absolute energies and mixing enthalpy for the binary alloys in this study.

**Methods**

Density functional theory calculations were performed with the Vienna *ab initio* simulation package (VASP) [18,19] and primarily with the Perdew–Burke–Ernzenhof generalized gradient approximation to model the exchange correlation energy [20]. For comparison, calculations were also done with the local density approximation (LDA) [21]. The projector augmented wave was used with an energy cutoff of 400 eV, and a 5x5x5 *k*-point sampling grid. The structures were relaxed until forces were smaller than 0.01 eV/Å. We used the special quasirandom structure (SQS) [22] approach with 64 atoms supercells for the $A_{1-x}B_x$ alloy compositions with x = 0.125, 0.250, 0.375, 0.500, 0.625, 0.750, 0.875. For consistency, also the pure binaries (x = 0.00 and 1.00) were calculated with the same supercell geometries and *k*-point settings.

For the UPMLIPs calculations, we used the final relaxed structure from the DFT calculations and relaxed it again using the corresponding UPMLIP. We report the energies and volumes from the ML-relaxed structures. For further analysis and methodology testing, we also performed DFT calculations and relaxations started from the obtained structures relaxed with MACE and compared energies, residual forces, and computational times with both fixed-ideal-lattice calculations and the fully relaxed DFT structures.

We focus our analysis of the enthalpy and energy results on the mixing enthalpies defined as:

$$\Delta H_{mix} = H(x) - xH(1) - (1-x)H(0) \qquad (1)$$

Absolute values of energies in both in UPMLIPs and DFT calculations depend heavily on specific methodologies, approximations, and numerical settings. The final VASP energy between calculations using the PBE and LDA functionals can differ several eV/atom, in part because these calculations use different pseudopotentials. The VASP energy should only be used to calculate relative energies between calculations using consistent settings. Mixing properties calculated within the same settings should thus reflect the real physical effects, regardless of the absolute energetics.

**Results and discussion**

Figure 1 shows the results for the total energy and mixing enthalpy at 0 GPa pressure for calculations done with DFT and the UPMLIPs. We show the volume and mixing volume in Figure S1 of the supplementary materials. All results are given per atom. Overall, the UPMLIPs exhibit a mixed performance over the studied alloys, in some cases, such as AgCu, all UPMLIPs show a good agreement with DFT, whereas in others they are in complete disagreement. For example, in AlZn all the UPMLIPs predict the opposite sign for the mixing energy. In general, none of the UPMLIPs performs the best, they all demonstrate a varying degree of success and failure across different systems. The lack of performance of UPMLIPs for formation energies in alloys may seem surprising when compared to the reported performance with surfaces [23]. However, if one only considers the total energies, it may appear that the UPMLIPs are in good agreement with the DFT results, as in the case for AlZn. However, for the mixing energy, this is no longer true, and we even see predictions with a mixing behavior opposite to the known correct predictions of DFT. Commonly, MLIPs can be trained below a MAE of 10 meV/atom for energy predictions [24], which is not sufficient for mixing energies, as one can see in our results. In DFT calculations, absolute energies are not meaningful, and relative energies are the important quantity. As such it is not important which pseudopotential is used in DFT calculations, for example the shift ion total energy for pure Cu in the Cu bearing compounds, as seen in Figure 1, comes from the DFT calculations and ML fit using different flavors of the Cu pseudopotential yet this has no consequence in the extracted physical quantities. Generally, while the relative errors between similar systems are far lower than the overall error of DFT, this behavior appears not to transfer so cleanly to UPMLIPs, at least not for mixing energies of alloys.

One can compare the differences between the different UPMLIPs and DFT with that of using different exchange-correlation functionals in DFT. Therefore, we have also included the results for DFT-LDA calculations for the mixing enthalpies. The LDA and PBE exchange-correlation functionals are constructed based on different principles, and frequently yield differences in results; e.g., LDA and PBE tend to respectively underestimate and overestimate lattice constants in cell shape relaxations. We do not include the LDA results in the total energy, since it would only shift the energy compared to PBE. Similarly, we do not include them in the volume results in the supplementary material, since LDA is known to give smaller volumes than PBE. For the mixing energies, LDA and PBE generally agree well, within 10 meV/atom. However, LDA has large issues in certain specific systems, such as the coupling of the magnetic state and volume in fcc-Fe [25], which is why LDA curves for AuFe and FePd are absent from the mixing enthalpy plots. In cases such as AlZn LDA and PBE provide almost indistinguishable mixing enthalpies. DFT typically has an error of approximately 20 meV compared to experiments, primarily due to systematic errors or noise, although mixing enthalpies benefit from error cancellation and tend to be small enough to correctly predict the right mixing trends.

A qualitative summary of the performance of the universal potentials in the current alloy theory application is shown in Table 1. A color system is used to assess the performance based on the mixing enthalpies, green being a quantitative and qualitative agreement, with differences being less than 10% or 10 meV/atom, yellow represents a qualitative agreement with general correct sign and shape but absolute values differing between 10 and 50%, and red marks a completely failed prediction. At the bottom of the table, we include the mean absolute error and the mean square error for the three universal potentials taken at the composition $x$=0.50. Table 1 shows that no UPMLIP gives consistently the best predictions.

In a recent work, Yu *et. al* [26] analyzed the reliability of UPMLIPs for structure optimization. They observed that CHGNET performs slightly better than MACE for certain chemistries, primarily when the composition includes transition metals. The architecture of CHGNET includes a coarse-grained notion of atomic charges in the training data using magnetic moments, allowing the model to distinguish multivalent elements. This inclusion of magnetic moments likely improves the generalizability of the potential to certain systems. The overall performance of the different UPMLIPs have been surveyed by Matbench [27], an automated leaderboard for benchmarking ML algorithms.

Table 1. Universal potentials performance for the 21 alloy systems considered in this study.

|      | DFT-LDA | MACE | CHGNET | M3GNET |
|------|---------|------|--------|--------|
| AgCu | green   | green | yellow | green |
| AgPt | green   | green | red    | red    |
| AlZn | green   | red   | red    | red    |
| AsSb | green   | red   | red    | red    |
| AuCu | 🟡 | 🟢 | 🟡 | 🟡 |
| AuFe | n/a | 🟡 | 🟡 | 🔴 |
| AuNi | 🟢 | 🟡 | 🟡 | 🟢 |
| AuPt | 🟢 | 🔴 | 🔴 | 🟡 |
| CrFe | 🟡 | 🔴 | 🔴 | 🟡 |
| CrMo | 🟢 | 🔴 | 🔴 | 🟢 |
| CrW | 🟢 | 🔴 | 🔴 | 🟢 |
| CuNi | 🟢 | 🟡 | 🔴 | 🔴 |
| CuRh | 🟢 | 🟡 | 🔴 | 🔴 |
| FePd | n/a | 🟡 | 🔴 | 🟡 |
| FeZn | 🟡 | 🔴 | 🔴 | 🔴 |
| IrPd | 🟢 | 🟡 | 🟢 | 🟡 |
| IrPt | 🟢 | 🟡 | 🟡 | 🟡 |
| IrRh | 🟢 | 🔴 | 🟢 | 🔴 |
| NiRh | 🟡 | 🔴 | 🔴 | 🔴 |
| PdPt | 🟢 | 🟢 | 🟢 | 🟡 |
| PdRh | 🟢 | 🟡 | 🟡 | 🟢 |
| MAE x=0.5 (meV/atom) | 0.009417 | 0.035850 | 0.041524 | 0.027697 |
| MSE x=0.5 (meV/atom) | 0.000135 | 0.002585 | 0.002839 | 0.001447 |

The striking difference in prediction between DFT and the universal potentials in the case of AlZn might originate from being results from different structures, meaning that the relaxation using the UPMLIPs could have found a different structure than the DFT relaxation. To investigate this possibility, we used the output structure from the MACE-UPMLIP relaxations as input for a static DFT calculation, followed by a relaxation to check if the structure would relax into a different optimized structure. As shown in Figure 2, this is not the case. The energies from PBE-DFT on the MACE-UPMLIP structure are practically identical to those from the fully relaxed PBE-DFT calculation. This means that the DFT and MACE-UPMLIP had relaxed into very similar structures, including local lattice relaxations. This agreement in structural relaxation suggests an avenue to accelerate our calculations in the current setting. Since the prediction of forces by the UPMLIPs is good enough to relax the structures, one could start by performing structural relaxations with UPMLIPs and then use the relaxed structure for a one-shot static DFT calculation to obtain the mixing energies. This approach will not yield energies as accurate as a full DFT workflow but accelerates possibly computationally expensive DFT relaxation steps. To test this approach, we calculated the energies for the four systems AgCu,

AlZn, AsSb, and AuNi with PBE-DFT on the PBE-DFT relaxed structure; on the MACE-relaxed structure; and on a fixed ideal lattice with the volume from Vegard's rule.

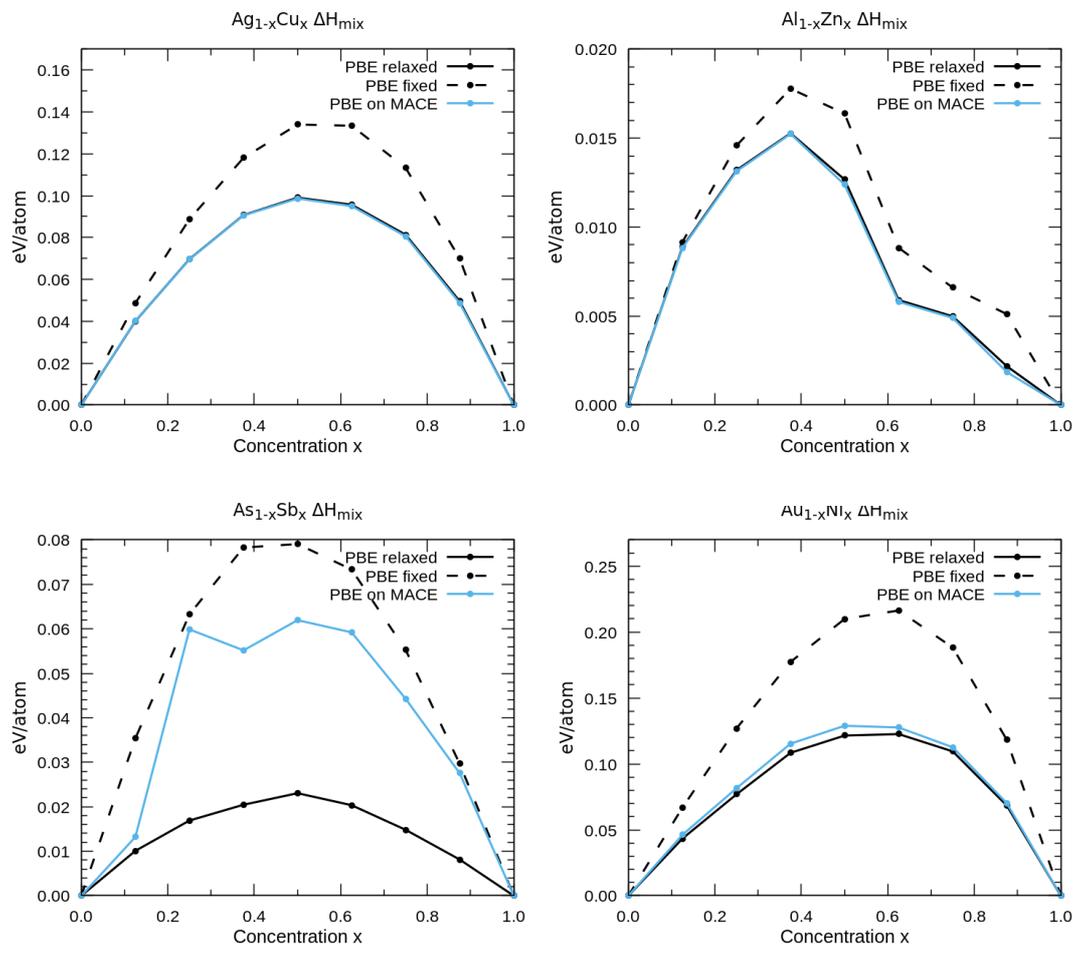

Figure 2. Comparison of results between the PBE-DFT relaxed structure, on the MACE-relaxed structure, and on a fixed ideal lattice with the volume from Vegard's rule.

As mentioned above, the works presenting the UPMLIPs have suggested that they can be used as a starting point to retrain the potentials for specific problems and material systems. To test this approach, we have chosen the AlZn system to verify if the correct mixing enthalpy is obtained with a model where the training data is supplemented by a few additional relevant systems. The MACE potential was retrained using the "large" foundation model as the base model, using float64 precision. Our training data is comprised of three starting structures with varying aluminum concentrations relative to zinc: 0.25, 0.50, and 0.75. We perturbed the initial structures to cover different local environments and local energies landscapes. We did not apply any corrections [28] to the energy in order to maintain compatibility with the MACE pre-trained model. A total of 34 structures were used to retrain the model for 50 epochs. Figure 3 shows the results of the retrained potentials, including the original MACE potential, a potential retrained with data generated with the same setting as the previous DFT simulations in this study, and a potential retrained with data generated from calculations using the same VASP settings, INCAR, KPOINTS, and pseudo potentials as used in the Materials Project [9]. The importance of having consistent settings when retraining potentials is highlighted by the fact that the expected positive mixing energy result is only achieved when the data used matches the settings of the pretrained potential.

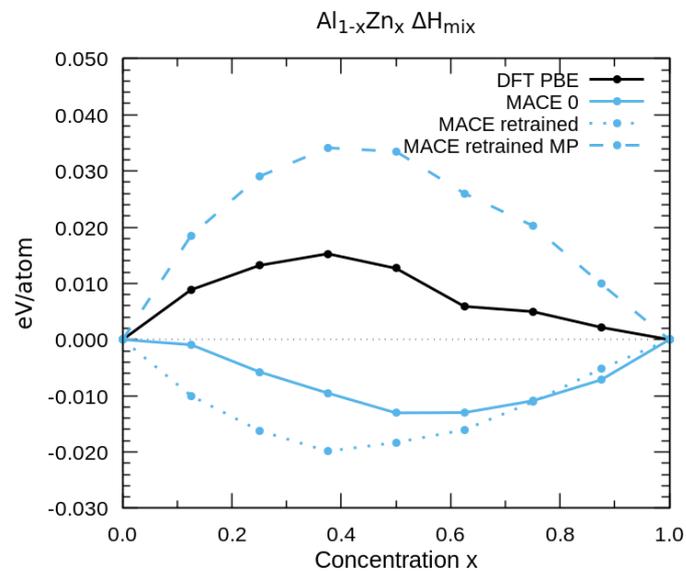

Figure 3. Results for the AlZn system after retraining the MACE potential. The result of the original MACE potential is labeled MACE 0. The potential retrained with data generated with the same setting as the previous DFT simulations in this study (which differ in the energy cutoff value) is labeled MACE retrained, and the retrained potential with the same INCAR settings, KPOINTS and pseudo potentials as the Materials Project is labeled MACE retrained MP.

**Conclusions**

The adoption of ML approaches offers an avenue to revolutionize materials research by enabling simulations for exciting and challenging new problems. These methods make it possible to study complex processes, such as cracking and fracture mechanisms with comparable accuracy to *ab-initio* methods. Recently developed universal potentials can, in principle, treat any material system in any problem setting, with their usage exemplified in complex simulation settings. Yet, we demonstrate that none of the universal potentials examined in the present work can be directly applied to predict mixing energies of isostructural binary alloys that agree with DFT results. The mixing enthalpy of alloys is a fundamental and important parameter for industrial applications. The impressive results reported for the UPMLIPs may make it tempting to use them as is, without further testing or retraining. Our work shows that this should not be done without careful consideration of accuracy. We stress the need to perform tests and, when needed, extend the training with data relevant for the targeted application. Specifically for mixing properties of alloys, UPMLIPs do not predict relative energies with sufficient accuracy even for use in a low accuracy pre-screening step for, e.g., a high-throughput screening effort. However, as suggested in the papers introducing these potentials, we indeed find that they can be used as starting points that can be adapted by further training for a particular problem. We have also found that the MACE potential can be used as a pre-relaxer for alloys, generally capable of reducing the DFT-simulation to a single static run.

**Acknowledgements**

B.A. acknowledges financial support from the Swedish Research Council (VR) through Grant No. 2019-05403 and 2023-05194, from the Swedish Government Strategic Research Area in Materials Science on Functional Materials at Linköping University (Faculty Grant SFOMatLiU No. 2009-00971). R.A. and A.P acknowledges support from the Swedish Research Council (VR) grant no. 2020-05402 and the Swedish e-Science Centre (SeRC). The computations were enabled by resources provided by the National Academic Infrastructure for Supercomputing in Sweden (NAISS) partially funded by the Swedish Research Council through grant agreement no. 2022-06725.

# References

[1] D. R. Bowler and T. Miyazaki, *J. Condens. Matter Phys*., 074207 (2010).
[2] T. Wen, C.Z. Wang, M.J. Kramer, Y. Sun, B. Ye, H. Wang, X. Liu, C. Zhang, F. Zhang, K.M. Ho, and N. Wang, *Phys. Rev. B* 100, 174101 (2019).
[3] A. V. Shapeev, *Multiscale Model. Sim*. 14, 1153 (2016).
[4] Y. Zuo, C. Chen, X. Li, Z. Deng, Y. Chen, J. Behler, G. Csányi, A.V. Shapeev, A.P. Thompson, M.A. Wood, and S.P. Ong, *J. Phys. Chem*. A 124, 731 (2020).
[5] T. W. Ko and S. P. Ong, *Nat. Comput. Sci*. 3, 998 (2023).
[6] C. Chen and S. P. Ong, *Nat. Comput. Sci*. 2, 718 (2022)..
[7] B. Deng, P. Zhong, K. Jun, J. Riebesell, K. Han, C. J. Bartel, and G. Ceder, *Nat. Mach. Intell.* 5, 1031 (2023).
[8] I. Batatia et al., *arXiv preprint* arXiv:2401.00096 (2023)
[9] A. Jain et al., *APL materials* 1 (2013).
[10] M. Hellenbrandt, *Crystallogr. Rev.*, *10*(1), 17–22 (2004).
[11] V.I. Anisimov, J. Zaanen, and O.K. Andersen Phys. Rev. B 44, 943 (1991).
[12] Deng, B., Choi, Y., Zhong, P., Riebesell, J., Anand, S., Li, Z., … & Ceder, G., *arXiv preprint arXiv:2405.07105* (2024).
[13] A. Wang, R. Kingsbury, M. McDermott, . et al., *Sci Rep* 11, 15496 (2021)
[14] C. Chen, S.P. Ong, MPF.2021.2.8. figshare. Dataset. https://doi.org/10.6084/m9.figshare.19470599.v3, (2022).
[15] D. Bowen, Materials Project Trajectory (MPtrj) Dataset. figshare. Dataset. https://doi.org/10.6084/m9.figshare.23713842.v2, (2023).
[16] R. Drautz, *Phys. Rev. B* 99, 014104 (2019).
[17] G. Dusson, M. Bachmayr, G. Csányi, R. Drautz, S. Etter, C. van Der Oord, and C. Ortner, *J. Comput. Phys.* 454, 110946 (2022).
[18] G. Kresse and J. Furthmüller, *Comput. Mater. Sci*. 6, 15 (1996).
[19] G. Kresse and J. Furthmüller, Phys. Rev. B 54, 11169 (1996).
[20] J. P. Perdew, K. Burke, and M. Ernzerhof, *Phys. Rev. Lett*. 77, 3865 (1996).
[21] P. E. Blöchl, *Phys. Rev. B* 50, 17953 (1994).
[22] A. Zunger, S. H. Wei, L. G. Ferreira, and J. E. Bernard, *Phys, Rev. Lett*. 65, 353 (1990).
[23] B. Focassio, L. P. M. Freitas, and G. R. Schleder, *arXiv preprint* arXiv:2403.04217 (2024).
[24] J. Riebesell, R. E. Goodall, A. Jain, P. Benner, K. A. Persson, and A. A. Lee, *arXiv preprint* arXiv:2308.14920 (2023).
[25] I. A. Abrikosov, A. E. Kissavos, F. Liot, B. Alling, S. Simak, O. Peil, and A. V. Ruban, *Phys. Rev. B* 76, 014434 (2007).
[26] H. Yu, M. Giantomassi, G. Materzanini, and G.-M. Rignanese, *arXiv preprint* arXiv:2403.05729 (2024).
[27] Dunn, A., Wang, Q., Ganose, A., Dopp, D., Jain, A, *npj Comput. Mater*. 6, 138 (2020).
[28] S. P. Ong *et al.*, *Comput. Mater. Sci*. 68, 314 (2013).